\documentclass[prl,aps,showpacs,twocolumn]{revtex4}
\usepackage{epsfig}
\begin{document}
\author{Li-Bin Fu}
\affiliation{Institute of Applied Physics and Computational
Mathematics, P.O. Box 8009
(28), 100088 Beijing, China, and \\
Max-Planck-Institute for the Physics of Complex systems,
N\"{o}thnitzer Str. 38, 01187 Dresden, Germany}
\title{General correlation functions of the Clauser-Horne-Shimony-Holt inequality for arbitrarily high-dimensional
systems}
\date{\today}

\begin{abstract}
We generalize the correlation functions of the
Clauser-Horne-Shimony-Holt (CHSH) inequality to arbitrarily
high-dimensional systems. Based on this generalization, we
construct the general CHSH inequality for bipartite quantum
systems of arbitrarily high dimensionality, which takes the same
simple form as CHSH inequality for two-dimension. This inequality
is optimal in the same sense as the CHSH inequality for two
dimensional systems, namely, the maximal amount by which the
inequality is violated consists with the maximal resistance to
noise. We also discuss the physical meaning and general definition
of the correlation functions. Furthermore, by giving another
specific set of the correlation functions with the same physical
meaning, we realize the inequality presented in [Phys. Rev. Lett.
{\bf 88, }040404 (2002)].
\end{abstract}

\pacs{03.65.Ud, 03.67.-a} \maketitle Since the foundations of
quantum mechanics were laid, one of the most remarkable aspects of
quantum mechanics is its predicted correlations. The quantum
correlations between outcomes of measurements performed on quantum
entangled states of systems composed of several parts have no
classical analog. Historically, this became known as the
Einstein-Podolsky-Rosen paradox \cite{epr} and was formulated in
terms of measurable quantities by Bell \cite{bell} as the nowadays
famous Bell inequalities. Subsequently, a vast amount of
literature has covered lots of aspects, ranging from philosophy to
experimental physics. One of the most common form of Bell
inequalities is known as the Clauser-Horne-Shimony-Holt (CHSH)
inequality \cite{CHSH} which is described in terms of correlation
functions by considering the correlations between measurements
performed on two entangled spin-$1/2$ particles.

Recently, two kinds of inequalities \cite{gisin,dag} have been
found that generalize the CHSH inequality to systems of higher
dimension. The authors of Ref. \cite{gisin} developed a new Bell
inequality, denoted here as CGLMP inequality, for arbitrarily
high-dimensional systems in terms of joint probabilities. Based on
this inequality, the authors gave the analytic description of
previous numerical results \cite{mzk}. For the two-dimensional
systems (systems composed of two spin-$1/2$ particles), this
inequality reduces to the familiar CHSH inequality. Alternative to
this inequality, the authors of Ref. \cite{dag} obtained an
inequality for three-dimensional systems in terms of correlation
functions. These two inequalities are equivalent for
three-dimensional systems \cite{acin,fu}.

In this letter, inspired by the previous efforts \cite{gisin,dag},
we generalize the correlation functions of the CHSH inequality for
bipartite two-dimensional systems to arbitrarily high-dimensional
systems. Then we construct a new Bell inequality for the
arbitrarily high dimensional system by using these correlation
functions. The new inequality is not only of the same form, but
also optimal in the same sense as the CHSH inequality, i.e., the
maximal amount by which the inequality is violated consists with
the maximal resistance to noise. Furthermore, we give a physical
interpretation of the correlation function, and discuss the
possible equivalent definitions of the correlation functions with
the same physical meaning. By employing a specific set of
correlation functions, we obtain the CGLMP inequality.

The scenario of the inequality involves two parties: Alice, can carry out
two possible measurements, $A_{1}$ or $A_{2}$, on one of the particles,
whereas the other party, Bob, can carry out two possible measurements, $%
B_{1} $ or $B_{2}$, on the other one. For the composed systems of $d$%
-dimensional parties (or bipartite systems of spin $S$ particles with the
relation $d=2S+1),$ each measurement may have $d$ possible outcomes: $%
A_{1},A_{2},B_{1},B_{2}=0,\ldots ,d-1$. The joint probabilities
are denoted by $P(A_{i},B_{j})$, which are required to satisfy the
normalization condition: $\sum_{m,n=0}^{d-1}P(A_{i}=m,B_{j}=n)=1$
.

The CHSH inequality \cite{CHSH} for two entangled spin-$1/2$ particles reads
\begin{equation}
<A_{1}B_{1}>+<A_{1}B_{2}>-<A_{2}B_{1}>+<A_{2}B_{2}>\leq 2,
\label{eeaa}
\end{equation}
where the functions $<A_{i}B_{j}>$ \ are the expectation values of products $%
A_{i}\otimes B_{j}$ measured on pairs, known as the correlation functions. $%
\ $The inequality will never be violated by a local hidden
variable theory, but will be maximally violated with the factor
$\sqrt{2}$ by quantum predications of an maximally entangled
state. On the other hand, the correlation functions can be
expressed in terms of joint probabilities by
\begin{equation}
<A_{i}B_{j}>=\sum_{m=0}^{1}\sum_{n=0}^{1}(-1)^{n+m}P(A_{i}=m,B_{j}=n).
\label{ab}
\end{equation}

More recently, the authors of Ref. \cite{dag} gave a CHSH-type inequality
for three-dimensional systems (for spin-$1$ particles), which reads
\begin{widetext}
\begin{equation}
I={\rm Re}[\bar{Q}_{11}+\bar{Q}_{12}-\bar{Q}_{21}+\bar{Q}_{22}]+\frac{1}{%
\sqrt{3}}{\rm Im}[\bar{Q}_{11}-\bar{Q}_{12}-\bar{Q}_{21}+\bar{Q}_{22}]\leq 2
\label{ch1}
\end{equation}
\end{widetext}
where the correlation functions $\bar{Q}_{ij}$ are defined as follows:
\begin{equation}
\bar{Q}_{ij}=\sum\limits_{m,n=0}^{2}\alpha ^{n+m}P(A_{i}=m,B_{j}=n),
\label{qq1}
\end{equation}
in which $\alpha =e^{i2\pi /3}.$ Let us reform it as follows.
Since the joint probabilities are real, we can simplify
(\ref{ch1}) to
\begin{equation}
I=Q_{11}+Q_{12}-Q_{21}+Q_{22}\leq 2.  \label{ch3}
\end{equation}
by defining $Q_{ij}=%
\mathop{\rm Re}%
[\bar{Q}_{ij}]+1/\sqrt{3}%
\mathop{\rm Im}%
[\bar{Q}_{ij}]$ for $i\geq j,$ and $Q_{12}=%
\mathop{\rm Re}%
[\bar{Q}_{12}]-1/\sqrt{3}%
\mathop{\rm Im}%
[\bar{Q}_{12}].$ Obviously, (\ref{ch3}) has the same form of the
CHSH inequality (\ref{eeaa}).\ Furthermore, from (\ref{qq1}), we
can prove that the new correlation functions $Q_{ij}$ can be
written in the following form:
\begin{equation}
Q_{ij}\equiv \frac{1}{S}\sum_{m=0}^{d-1}%
\sum_{n=0}^{d-1}f^{ij}(m,n)P(A_{i}=m,B_{j}=n),  \label{qq}
\end{equation}
in which $S=1$, the spin of the particle for the $3$-dimensional
system, $ f^{ij}(m,n)=S-M(\varepsilon (i-j)(m+n),d)$, and
$\varepsilon (x)$ is the sign function: $\varepsilon (x)=\left\{
\begin{array}{cc}
1 & x\geq 0 \\
-1 & x<0
\end{array}
\right..  $$M(x,d)$ is defined as follows: $M(x,d)=(x\mbox{ mod
d})$ and $0\leq M(x,d)\leq d-1$. Comparing (\ref{qq}) with
(\ref{ab}), we find that the correlation functions for two
entangled spin-$1/2$
particles can also be expressed with (\ref{qq}) by substituting $S=1/2$ and $%
d=2$ correspondingly.

Obviously, the formula (\ref{qq}) generalizes the correlation
function to arbitrarily dimensional systems.

At the same time, we assume the CHSH inequality expression for
arbitrarily dimensional systems takes the same form as the CHSH
inequality for two-dimensional systems, namely
\begin{equation}
I_{d}=Q_{11}+Q_{12}-Q_{21}+Q_{22}.  \label{bd}
\end{equation}
$I_{d}$ is upper bounded by $4.$ This follows immediately from the
fact that the extreme values of $Q_{ij}$ are $\pm 1.$ However,
these four functions are strongly correlated, so $I_{d}$ can never
reach this value. In a local
hidden variable theory, only three of the four pairs of operators: $(A_{1},B_{1}),$ $%
(A_{1},B_{2}),$ $\left( A_{2},B_{1}\right) $ and $\left(
A_{2},B_{2}\right) , $  can be freely chosen, the last one is
constrained. We can prove the maximum value of $I_{d}$ for local
hidden variable theories is $2$, i.e., $I_{d}\leq 2.$

The proof consists of enumerating all the possible relations
between pairs of operators allowed by the local hidden variable
theory. Defining $r_{11}\equiv A_{1}+B_{1},$ $r_{12}\equiv A_{1}+B_{2},$ $%
r_{21}\equiv A_{2}+B_{1}$ and $r_{22}\equiv A_{2}+B_{2}.$
Obviously, they obey the constraint
\begin{equation}
r_{11}+r_{22}=r_{12}+r_{21}.  \label{ddaa}
\end{equation}
The correlation functions (\ref{qq}) for a given choice of $r_{11},$ $%
r_{12}, $ $r_{21}$ and $r_{22}$ are, $Q_{11}=g_{1}(r_{11}),$ $%
Q_{12}=g_{2}(r_{12}),$ $Q_{21}=g_{1}(r_{21})$ and $Q_{22}=g_{1}(r_{22}),$%
where $g_{1,2}\left( x\right) $ are given by
\begin{eqnarray}
g_{1}(x) &=&\frac{S-M(x,d)}{S},\text{ }  \nonumber \\
g_{2}(x) &=&\frac{M(x,d)-S-1}{S}.  \label{gg}
\end{eqnarray}
Then we immediately have
\begin{equation}
I_{d}=\frac{M(r_{12},d)+M(r_{21},d)-M(r_{11},d)-M(r_{22},d)-1}{S}
\label{bdc}
\end{equation}
Now, we consider different cases according to the values of $%
r_{11},r_{12},r_{21},$ and $r_{22}.$

Case 1. Both $r_{11}$ and $r_{22}$ are less than $d$. From
(\ref{ddaa}), there are two cases for the rest: (i) none of
$r_{12}$ and $r_{21}$ is larger than $d$, (ii) one of them is
larger than $d.$ Then from (\ref{bdc}),
if none of $r_{12}$ and $r_{21}$ is larger than $d$, we get $%
I_{d}=[r_{12}+r_{21}-(r_{11}+r_{22}-d)-1]/S=2$ (keeping in mind
$d=2S+1);$ if one of $r_{12}$ and $r_{21}$ is larger than $d,$
then $I_{d}=-1/S.$

Case 2. One of $r_{11}$ and $r_{22}$ is larger than $d$. There are
three cases for the rest (i) none of $r_{12}$ and $r_{21}$ is
larger than $d$, (ii) one of them is larger than $d$, and (iii)
both $r_{12}$ and $r_{21}$
are larger than $d.$ On can find that there are three possible results for $%
I_{d}$: $I_{d}=2$, $I_{d}=-1/S,$ and $I_{d}=-2(S+1)/S.$

Case 3. Both $r_{11}$ and $r_{22}$ are larger than $d.$ Then
(\ref{ddaa}) implies: (i) both $r_{12}$ and $r_{21}$ are also
larger than $d$, (ii) one
of them is larger than $d.$ From (\ref{bdc}), one finds $I_{d}$ is either $%
I_{d}=-1/S$ or $I_{d}=2$.

Thus, for all possible choices of $r_{ij}$, $I_{d}$ $\leq 2$ for
local realism (Note that for $d=2$, not all the possibilities
enumerated above can occur. One can prove that the only possible
values are $I_{2}=\pm 2.$). Here, we must point out that the proof
is also valid for non-deterministic local hidden variable theories
for the convexity of the correlation polytope .

Let us now consider the maximum value that can be attained for the
Bell expression $I_{d}$ for quantum measurements on an entangled
quantum state.
For the maximally entangled state of two $d$-dimensional systems $\psi =%
\frac{1}{\sqrt{d}}\sum_{j=0}^{d-1}\left| l\right\rangle _{A}\left|
l\right\rangle _{B},$ we therefore first recall the optimal
measurements
performed on such a state described in \cite{gisin,mzk}. Let the operators $%
A_{i},$ $i=1,2,$ measured by Alice and $B_{j},$ $j=1,2,$ measured
by Bob, have the non-degenerate eigenvectors
\begin{eqnarray}
\left| m\right\rangle _{A_{i}}
&=&\frac{1}{\sqrt{d}}\sum_{l=0}^{d-1}\exp \left( i\frac{2\pi
}{d}l(m+\alpha _{i})\right) \left| l\right\rangle _{A},
\nonumber \\
\left| n\right\rangle _{B_{j}}
&=&\frac{1}{\sqrt{d}}\sum_{l=0}^{d-1}\exp \left( i\frac{2\pi
}{d}l(n+\beta _{j})\right) \left| l\right\rangle _{B}, \label{mn}
\end{eqnarray}
where $\alpha _{1}=0,$ $\alpha _{2}=1/2,$ $\beta _{1}=1/4,$ and
$\beta _{2}=-1/4.$ Thus the joint probabilities are \cite{gisin}
\begin{equation}
P_{QM}(A_{i}=m,B_{j}=n)=\frac{1}{2d^{3}\sin ^{2}[\pi (m+n+\alpha
_{i}+\beta _{j})/d]}.  \label{pqm}
\end{equation}
These joint probabilities have several symmetries. First of all we
can have the relation
\[
P_{QM}(A_{i}=m,B_{j}=n)=P_{QM}(A_{i}=m\pm c,B_{j}=n\mp c)
\]
for any integer $c.$

For convenience, in the following we use the symbol $\doteq $ to
denote equality modulus $d.$ Let us define the probabilities
$P(A_{i}\pm B_{j}\doteq k)$ by
\begin{equation}
P(A_{i}\pm B_{j}\doteq k)=\sum_{m=0}^{d-1}P(A_{i}=m,B_{j}=(k\mp
m\mbox{ mod d})).  \label{pnp}
\end{equation}
Then, from (\ref{pqm}) we have
\begin{equation}
P_{QM}(A_{1}+B_{1}\doteq c)=dP_{QM}(A_{1}=c,B_{1}=0).  \label{pap}
\end{equation}

Let us define $S_{z}=S-M(\varepsilon (i-j)(m+n),d).$ One can prove that $%
S_{z}=-S,-S+1,\cdots ,S-1,S$ for all the possible values of $m$
and $n$. By denoting $q(S_{z})=P_{QM}(A_{1}+B_{1}\doteq
S-S_{z})=\frac{1}{2d^{2}\sin ^{2}[\pi (S-S_{z}+1/4)/d]},$ and
using the above formula, we can prove:
$Q_{11}=Q_{12}=Q_{22}=-Q_{21}=Q_{d},$ where
\begin{equation}
Q_{d}=\frac{1}{S}\sum_{S_{z}=-S}^{S}S_{z}q(S_{z}).  \label{qd}
\end{equation}
Then, we can obtain the quantum prediction of Bell expression for
the maximally entangled state $I_{d}(QM),$ namely
\begin{equation}
I_{d}(QM)=4Q_{d}.  \label{gsin}
\end{equation}
One can prove that this result is the same as that obtained in
\cite{gisin}, and it consists with the numerical work in
\cite{mzk}. So, the measurements defined by (\ref{mn}) are optimal
for the maximally entangled states and $I_{d}(QM)$ is the
strongest violation of Bell expression $I_{d}$ for the maximally
entangled states of bipartite $d$-dimensional systems. From the
numerical work of \cite{mzk} and analytical result of
\cite{gisin}, we conclude that the general CHSH inequality
(\ref{bd}) is optimal in the same sense as the CHSH inequality is
optimal for two dimensional systems.

At the same time, we obtain the optimal correlation matrix $\widehat{Q}%
=\{Q_{ij}\}$ for the maximally entangled state for $d$-dimensional
systems, which is $\widehat{Q}=Q_{d}\left(
\begin{array}{cc}
1 & 1 \\
-1 & 1
\end{array}
\right) .$This matrix is well-known for $d=2$ (
$Q_{2}=\frac{\sqrt{2}}{2}$).

In fact, one can formulate other versions of Bell inequality for a
given experimental setup. In the interesting paper \cite{pit},
Pitowsky and Svozil have presented the general method for the
derivation of all Bell inequalities for each given experimental
setup, in which two specific cases have also been discussed.
Actually, the inequality suggested in Ref. \cite {dag} was
obtained by searching the optimal inequality on the correlation
polytope corresponding to the unbiased six-port beam splitters
measurements.

The CHSH inequality for two dimensional system is the most popular
form of Bell inequality amongst physicists, due to its simplicity
and optimality. Another important reason is that it demonstrates
the nature of quantum correlations clearly. As shown in
(\ref{eeaa}), the correlation functions of CHSH inequality for two
dimensional system have explicit physical meaning,
namely, the correlation functions\ are the expectation values of products $%
A_{i}\otimes B_{j}$ measured on pairs. Obviously, the general
correlation functions (for arbitrary dimensionality) can not be
understood in such a sense. Nevertheless, the correlation
functions, in general, imply some kind of correlation between the
measurements on pairs.

Indeed, the normalization condition of joint probabilities can be
rewritten as $\sum_{k=0}^{d-1}P(A_{i}+B_{j}\doteq k)=1.$
Naturally, the $d$ real numbers $P(A_{i}+B_{j}\doteq k)$
$(k=0,1,\cdots ,d-1)$ can be regarded as the probabilities of
eigenvalues $S_{z}$ ($S_{z}=-S,-S+1,\cdots ,S)$ of a spin $S$
system, $\widetilde{P}(S_{z})$,
 under the relations:
\begin{equation}
\widetilde{P}(S_{z})=P(A+B\doteq S-S_{z}),  \label{ppq}
\end{equation}

We define the correlation function of the two measurements $A$ and
$B$ as follows:
\begin{equation}
C(A,B)=\sum_{S_{z}=-S}^{S}S_{z}\widetilde{P}(S_{z})=\left\langle
S_{z}\right\rangle _{AB}.  \label{cp}
\end{equation}
The meaning of the above formula is apparent, i.e., the
correlation of the two measurements $A$ and $B$ can be interpreted
as average of spin projection for the imaginary system with spin
$S$ defined by (\ref{ppq}).

In analogy, we can imagine another system which is dual to the
former one by
\begin{equation}
\widetilde{P}^{-}(S_{z})=P(A+B\doteq -(S-S_{z}))  \label{pnq}
\end{equation}
Consequently, we have another correlation function. To distinguish
these two, we substitute $\widetilde{P}^{+}(S_{z})$ for
$\widetilde{P}(S_{z})$ in Eq. (\ref{ppq}). Then the two kinds of
correlation functions can be labelled as: $C^{\pm
}(A,B)=\sum_{S_{z}=-S}^{S}S_{z}\widetilde{P}^{\pm
}(S_{z})=\left\langle S_{z}\right\rangle _{AB}^{\pm }$ .

It is not difficulty to prove that the CHSH inequality can be
rewritten by this kind of correlation functions as
\begin{equation}
\frac{1}{S}[%
C^{+}(A_{1},B_{1})+C^{-}(A_{1},B_{2})-C^{+}(A_{2},A_{1})+C^{+}(A_{2},B_{2})%
]\leq 2.  \label{ree}
\end{equation}
Comparing with (\ref{bd}), we have $Q_{11}=C^{+}(A_{1},B_{1})/S,$ $%
Q_{21}=C^{+}(A_{2},B_{1})/S,$ $Q_{22}=C^{+}(A_{2},B_{2})/S,$ and $%
Q_{12}=C^{-}(A_{1},B_{2})/S$

In fact, the labels of $d$ possible outcomes for each side ($A$
and $B)$ are arbitrary. So, in general, if we have a mapping: $g:$
$(A,B)\rightharpoonup (0,1,\cdots ,d-1),$ namely, $g(A,B)=c$
$(c=0,1,\cdots ,d-1),$ which is an one to one mapping for fixed
value of $A$ (or $B$). Let $P(g(A,B)=k)$ be the sum of all the
joint probabilities $P(A,B)$ which satisfy $g(A,B)=k.$ We can
define the correlation functions of the pair measurements $A$ and
$B$ as $C^{\pm }(A,B)=\sum_{S_{z}=-S}^{S}S_{z}\widetilde{P}^{\pm
}(S_{z})=\left\langle S_{z}\right\rangle _{AB}^{\pm }$ with $\widetilde{P}%
^{\pm }(S_{z})=P(g(A,B)\doteq \pm (S-S_{z})).$ Then from
(\ref{ree}), a CHSH type inequality can be established by
employing these correlation functions.

Especially, let us consider $g(A,B)=(A-B)\mbox{ mod d}.$ Defining $%
\widetilde{P}^{\pm }(S_{z})=P(A-B\doteq \pm (S-S_{z}))$, one can
prove that the sum $\sum_{S_{z}=-S}^{S}S_{z}\widetilde{P}^{\pm
}(S_{z})$ can be split into two parts as:
$\sum_{S_{z}=S_{0}}^{S}S_{z}[\widetilde{P}^{\pm
}(S_{z})-\widetilde{P}^{\pm }(-S_{z})],$ where $S_{0}=1/2$ for
even dimension (fermions) or $S_{0}=1$ for odd dimension (bosons).
Then from (\ref {ree}) and (\ref{pnp}), and letting $k=S-S_{z},$
we can get
\begin{eqnarray}
Q_{ij} &\equiv
&\sum_{k=0}^{[\frac{d}{2}]-1}(1-\frac{k}{S})[P\left(
A_{i}-B_{j}\doteq k\varepsilon (i-j)\right)   \nonumber \\
&&-P\left( A_{i}-B_{j}\doteq (-k-1)\varepsilon (i-j)\right) ],
\label{qrrr}
\end{eqnarray}
in which $[d/2]$ denotes the integer part of $d/2$ and we have
used the
formula $2S-k\doteq -k-1.$ Then from the expression of (%
\ref{bd}), we can obtain the CGLMP inequality \cite{gisin}. So,
the CGLMP inequality can be converted into the standard form of
CHSH inequality for arbitrarily high dimensionality by introducing
the general correlation functions.

From the above discussion, we can know that, for CHSH inequality,
the correlation between pair of measurements is described by
correlation functions. The correlation function has specific
physical meaning, namely, each pair of measurements can be casted
into the spin projection of an imaginary spin $S$ system, and the
correlation function is the expectation value of the spin
projection of the imaginary system. Especially, for the bipartite
system composed by two spin $1/2$ particles, the correlation
functions can be expressed by the expectation values of products
$A_{i}\otimes B_{j}$ measured on pairs. We do not know so far, for
the arbitrarily high-dimensional system, if the correlations
functions can be expressed by expectation values of products of
some kinds of general measurements of pairs. However, we think the
correlation functions of arbitrary dimensionality are worth to
study further .

In summary, we have constructed a Bell inequality for arbitrarily
high-dimensional systems by generalizing the correlation functions
of the CHSH inequality for bipartite two-dimensional systems to
arbitrarily high-dimensional systems. The main features of the
present work are two-fold: (i) we present the general correlation
functions for arbitrarily high-dimensional systems and establish
the general CHSH inequality by employing these correlation
functions. The general CHSH inequality is of the same form and
optimal in the same sense as the CHSH inequality for two
dimensional systems; (ii) we discuss the physical meaning of the
correlation functions and give a general description of the
correlation functions. The facts that the Bell inequality and as
well as the correlation functions have the unified forms and
physical meaning for arbitrarily dimensionality, suggest that the
quantum correlations should have some common properties for
arbitrary dimensionality, which will be useful for discussing
entanglement of systems of higher dimensionality as they have had
for two-dimensional systems \cite{gs}. Furthermore, the general
description of correlation functions make it possible for us to
construct some alternative CHSH type inequalities which may be
convenient to realized in experimental setups for higher
dimensional systems. So, we hope that the general CHSH inequality
and the general correlation functions presented here will draw
much more attention of physicists on studying Bell inequality and
entanglement of systems of large dimensionality.

The author is indebted to Dr. J.L. Chen for sharing his most
recent works on the CHSH inequalities for qutrits. We also thanks
Professor J.M. Rost, A. Buchleitner, and S.G. Chen for valuable
discussions. This work was supported by the 973 Project of China
and Science and Technology Funds of CAEP, and partly by the
Alexander von Humblodt Foundation. The author thanks the MPIPKS
for hosting during various stages of this work.

\end{document}